\documentclass[sigconf]{acmart}

\AtBeginDocument{%
  }

\copyrightyear{2025}
\acmYear{2025}
\setcopyright{cc}
\setcctype{by}
\acmConference[SoCC '25]{ACM Symposium on Cloud Computing}{November 19--21, 2025}{Online, USA}
\acmBooktitle{ACM Symposium on Cloud Computing (SoCC '25), November 19--21, 2025, Online, USA}\acmDOI{10.1145/3772052.3772270}
\acmISBN{979-8-4007-2276-9/2025/11}

\usepackage[usenames,dvipsnames]{xcolor,colortbl}
\usepackage{amsmath}
\usepackage[utf8]{inputenc}
\usepackage{amsmath,amsthm}
\usepackage{MnSymbol}
\usepackage{cleveref}
\usepackage{listings, fancyvrb, multicol}
\usepackage{enumitem}
\usepackage{float}
\usepackage{tikz-cd}
\usepackage{microtype}
\usepackage{listings}
\usepackage{comment}
\usepackage{xspace}
\usepackage{graphicx, graphics}
\usepackage{subcaption}
\usepackage{svg}
\usepackage{balance}
\usepackage{etoolbox}
\newcommand{\new}[1]{{#1}}

\begin{document}
\title{Understanding GPU Resource Interference One Level Deeper}

\author{Paul Elvinger}
\orcid{0009-0000-6025-844X}
\affiliation{%
  \institution{ETH Zurich}
  \country{Switzerland}
}

\author{Foteini Strati}
\orcid{0000-0003-3364-2109}
\affiliation{%
  \institution{ETH Zurich}
  \country{Switzerland}
}

\author{Natalie Enright Jerger}
\orcid{0000-0002-0526-2080}
\affiliation{%
  \institution{University of Toronto}
  \country{Canada}
}

\author{Ana Klimovic}
\orcid{0000-0001-8559-0529}
\affiliation{%
  \institution{ETH Zurich}
  \country{Switzerland}
}

\renewcommand{\shortauthors}{Elvinger et al.}

\settopmatter{authorsperrow=4}

\begin{abstract}

GPUs are vastly underutilized, even when running resource-intensive AI applications, as GPU kernels within each job have diverse resource profiles that may saturate some parts of a device while often leaving other parts idle. Colocating applications is known to improve GPU utilization, but is not common practice as it becomes difficult to provide predictable performance due to workload interference. Providing predictable performance guarantees requires a deep understanding of how applications contend for shared GPU resources such as block schedulers, compute units, L1/L2 caches, and memory bandwidth. We study the key types of GPU resource interference and develop a methodology to quantify a workload's sensitivity to each type. We discuss how this methodology can serve as the foundation for GPU schedulers that enforce strict performance guarantees and how application developers can design GPU kernels with colocation in mind to improve efficiency.

\end{abstract}


\maketitle 

\section{Introduction}

Graphics Processing Units (GPUs) are widely used in AI training and inference to maximize performance per Watt. The power needs of AI workloads now comprise a significant percentage of datacenter power~\cite{stojkovic2024dynamollm, Patel24power, samsi23wordswatts} and contribute significant cost, since GPU hardware is power hungry and expensive. To minimize total cost of ownership and make optimal use of the limited power budget, cloud providers need to operate GPU clusters at high utilization. Yet many recent studies show that GPUs are vastly underutilized~\cite{Xi1o2020Antman, zhao2023muxflow, Weng22mlaas, Gujarati2020Clockwork, Mingcong22Reef, Ng2023Paella, Strati24Orion, Gao24utilization}. 
Even when serving multi-billion-parameter LLMs with large batch sizes, some GPU components may remain idle as resource requirements vary across compute vs. memory intensive phases of a job~\cite{zhu2024nanoflow, kamath2024podattention}. For example, Microsoft reports less than 10\% compute utilization during the memory-bound decoding phase of the Llama3-8B model on A100 GPUs~\cite{kamath2024podattention}. 
GPUs can also be underutilized due to small batch sizes, communication, data preprocessing bottlenecks, and checkpointing~\cite{graur22cachew, zhang20network, Gao24utilization}.

GPU schedulers aim to improve utilization by colocating workloads. 
Temporal-sharing schedulers~\cite{Gujarati2020Clockwork, Xiao18Gandiva, Yu19Salus, Xi1o2020Antman, Wu23TGS}
execute one workload at a time to avoid resource interference, but fail to address single-workload GPU underutilization, and can cause severe queuing delays~\cite{Strati24Orion}. 
In contrast, spatial sharing systems~\cite{Sudipta24Usher, Strati24Orion, Mingcong22Reef, Xu2022iGniter, zhang2024missile, Lim21Zico}
allow concurrent workload execution, and propose strategies to minimize interference. However, existing spatial sharing systems do not provide reliable performance guarantees. Most systems rely on limited metrics to evaluate GPU resource utilization and define colocation strategies. 
As we show in \S\ref{sec:issues-of-existing-estimations}, state-of-the-art systems such as Orion~\cite{Strati24Orion} and Usher~\cite{Sudipta24Usher} oversimplify GPU utilization and interference modeling, resulting in colocation decisions that can significantly degrade application performance. This prevents users from applying such colocation mechanisms in practice.

Analogous to how CPU schedulers make informed scheduling decisions by understanding how workloads interfere on CPU resources~\cite{Delimitrou14Quasar, Delimitrou13Paragon}, designing efficient GPU schedulers that spatially share resources while providing performance guarantees requires a deep understanding of GPU utilization and interference.
In this paper, we characterize how GPU workloads (e.g., LLM decode) utilize and contend for resources within a GPU and discuss the implications for designing GPU kernels and scheduling systems. 
We study key sources of interference that arise from sharing the multifaceted components of GPUs, including streaming multiprocessors, warp schedulers, computation cores, high-bandwidth memory, caches, and shared memory. \new{While some of these interference sources (memory bandwidth, caches) are well-known from CPUs, they remain largely understudied in modern GPUs.} We propose techniques to assess the sensitivity of GPU kernels and workloads to contention for each type of resource, using a combination of GPU profiling tools and a customizable suite of GPU microbenchmarks. 
\new{Our microbenchmarks are available at \href{https://github.com/eth-easl/gpu-util-interference}{https://github.com/eth-easl/gpu-util-interference} and our LLM profiling scripts are available at \href{https://github.com/eth-easl/vllm_profile}{https://github.com/eth-easl/vllm\_profile}.}

We draw several key takeaways from our analysis. First, scheduling decisions should be made at fine (per-kernel) granularity, to avoid head-of-line blocking at the GPU block scheduler. Second, allocating kernels to separate Streaming Multiprocessors (SMs) helps eliminate some sources of interference, but contention for shared resources like L2 cache and memory bandwidth can still cause significant slowdowns. Third, sharing SMs between kernels can be beneficial, though it is subject to multiple sources of interference, such as shared memory bandwidth, warp scheduling, and compute pipelines' contention. Finally, we observe that trading off per-kernel marginal performance improvements for higher colocation opportunities can significantly benefit GPU efficiency and cost. Using these insights, we describe ways for designing GPU schedulers that effectively colocate applications to reduce cost while providing strict performance guarantees. We also discuss implications for GPU application programmers, such as how to design GPU kernels to be more suitable for colocation to improve efficiency. 

\section{Background}\label{sec:background_motiv}

We describe the internal architecture of GPUs, introducing terminology and utilization metrics that we will refer to later.

\subsection{GPU Hardware Overview}\label{sec:background:gpu-arch}

\begin{figure}
    \centering
    \includegraphics[width=\linewidth]
    {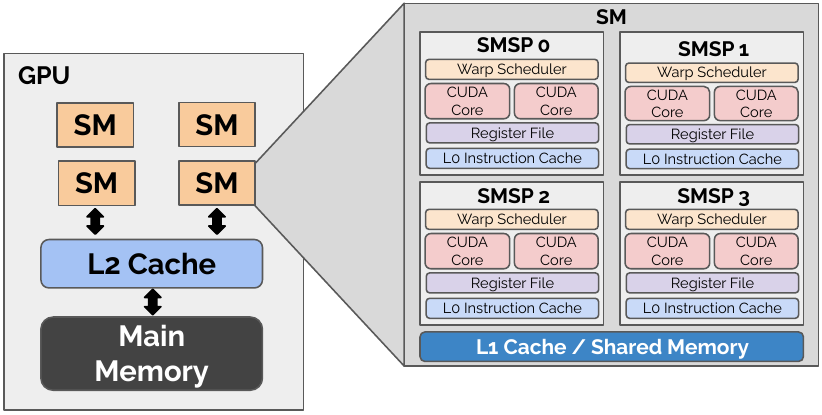}
    \caption{Simplified diagram of an NVIDIA GPU (based on an H100), focusing on a Streaming Multiprocessor (SM).}
\label{fig:gpu_arch}
\vspace{-11pt}
\end{figure}

\textbf{GPU architecture: } Figure \ref{fig:gpu_arch} depicts a modern GPU.\footnote{We focus on NVIDIA GPUs and terminology. AMD GPUs follow similar architecture~\cite{amd-description}, and provide tools such as Omniperf to get insights about kernels' execution~\cite{omniperf}.} 
GPUs consist of multiple clusters of Streaming Multiprocessors (SMs).
Each SM consists of subpartitions (SMSP) that contain a warp scheduler (which can schedule 32 threads/cycle), an L0 instruction cache, a register file, and various compute units for different data types (e.g., int32, fp32) and different operations, referred to as \textit{pipelines} (e.g., tensor cores)~\cite{ncu-metrics-decoder}. 
The mapping of pipelines to compute units is hidden from users, though there have been attempts to reverse engineer it~\cite{posthardware, postmapping, postpipelinesfp16, postsmsp}. The GPU contains a main memory shared by all SMs, accessed through a two-level on-chip cache hierarchy. The L2 cache is shared across all SMs, while each SM has a private memory space combining L1 cache and shared memory. The allocation between L1 cache and shared memory can be manually configured by the user~\cite{cuda-shared-mem-conf}.

\textbf{Threads, Blocks, and Warps.} GPU programs consist of CUDA kernels, executed by one or more GPU threads~\cite{cuda}. From a software perspective, GPU threads are grouped into \textit{blocks}, which are arranged in a \textit{grid}. Each thread has access to a set of registers and shared memory for its whole lifetime. Threads in a block communicate through shared memory and synchronize using barriers or other atomic operations. At the hardware level, the GPU executes threads in groups called \textit{warps}, typically consisting of 32 threads. Each kernel is associated with a \textit{CUDA stream}, which defines sequential execution of operations. If enough resources are available, kernels launched from multiple streams can execute concurrently. 

\textbf{GPU Scheduling}: NVIDIA GPUs schedule kernels at multiple levels.
First, the thread block scheduler maps thread blocks to SMs. A block remains on an SM until all its threads complete execution. Scheduling is constrained by SM resource limits (max number of blocks, threads, registers, and shared memory) that depend on the GPU architecture.
A block is scheduled on an SM only if that SM has enough resources left to accommodate all threads of that block. Once a block is scheduled on an SM, its warps are assigned to one of the subpartitions and are considered active.  
Each clock cycle, the warp scheduler in each subpartition chooses one eligible warp and schedules one or more instructions from that warp. 

\subsection{GPU Utilization Metrics}\label{sec:background:metrics} 

Most GPU schedulers rely on GPU utilization metrics to make colocation decisions.  There are many different GPU utilization metrics due to the multi-faceted nature of GPU resources. We outline popular metrics used in prior works below and metrics that will be used throughout our analysis. Most metrics can be reported by NVIDIA tools such as Nsight Compute (NCU)~\cite{ncu}.

\textbf{GPU utilization from \texttt{nvidia-smi}/NVML}~\cite{nvidiasmi, nvml} depicts the percentage of time at least one kernel is active on a GPU,  without revealing how well the kernels utilize the various GPU resources. This metric is used by various works~\cite{he2024uellm, Wu23TGS, Bhasi24serverless, Yeung20utilpred, Yeung22Horus}.

\textbf{SM utilization} refers to the number of SMs needed by a kernel, taking into account the kernel's grid size, block size, registers/thread, shared memory/thread, and the respective SM resource limits of a GPU. 
Orion~\cite{Strati24Orion} and REEF~\cite{Mingcong22Reef} use this metric.

\textbf{Arithmetic intensity} refers to the ratio of floating-point operations to total data movement, and can be found using NCU's roofline model~\cite{ncu-roofline}. Orion~\cite{Strati24Orion} uses this metric to classify kernels as compute-bound or memory-bound.

\textbf{Achieved occupancy} measures how many active warps exist per SM per clock cycle on average~\cite{achieved-occupancy, achievedoccblog} and is obtained by NCU, using the \texttt{sm\_\_warps\_active.avg.pct\_of\_peak\_sustained} \\ \texttt{\_active} metric. It is used by Usher~\cite{Sudipta24Usher} to estimate the compute requirement of a kernel. 

\textbf{Pipe utilization} measured by the NCU metric \texttt{sm\_\_inst\_executed} \\
\texttt{\_pipe\_*.avg.pct\_of\_peak\_sustained\_active} (* indicates the pipeline, e.g. \texttt{fma}, \texttt{tensor}) expresses how effectively a pipeline is used (when executing at least one warp) relative to its peak performance.

\textbf{Issued instructions per cycle (IPC):} The NCU metric
\texttt{sm\_\_inst} \\ \texttt{\_issued.avg.per\_cycle\_active} (also called \texttt{IPC}~\cite{ipcblog}) represents the average number of warp instructions issued per cycle per SM.
Our GPUs have 4 subpartitions per SM, each capable of issuing one warp instruction per cycle, i.e., the maximum IPC per SM is 4.
 
\textbf{L2 cache throughput and hit rate} The metrics \\ \texttt{lts\_\_throughput.avg.pct\_of\_peak\_sustained\_elapsed} and \\
\texttt{lts\_\_t\_sector\_hit\_rate.pct} measure the average L2 cache throughput and hit rate. 

\textbf{Shared Memory instruction bandwidth:} The NCU metric \texttt{l1tex\_\_data\_pipe\_lsu\_wavefronts\_mem\_shared.sum.pct\_of} \\
\texttt{\_peak\_sustained\_elapsed} measures the performance of shared memory load/store wavefronts\footnote{A \textit{wavefront} is the maximum unit that can pass through a pipeline stage per cycle~\cite{ncu-metrics-decoder}.} processed through the L1 data pipe. It provides insights into how effectively the available shared memory bandwidth is used.

\new{\textbf{Memory bandwidth utilization:} The NCU metric \texttt{gpu\_\_dram} \\ \texttt{\_throughput.avg.pct\_of\_peak\_sustained\_elapsed} measures the utilization of the memory bandwidth.}
\section{Pitfalls of GPU schedulers}
\label{sec:issues-of-existing-estimations}

Many existing GPU schedulers aim to improve utilization by colocating workloads, with temporal and/or spatial sharing. We focus on spatial sharing schedulers as they allow concurrent workload execution and hence require strategies to minimize interference.

We find that state-of-the-art GPU schedulers consider only a \textit{subset} of GPU utilization metrics, resulting in unreliable performance guarantees. We present common pitfalls below:

\textbf{Pitfall 1: Relying on a single utilization metric.} Most schedulers rely on a single metric to assess GPU utilization and make colocation decisions. For example, Usher~\cite{Sudipta24Usher} relies on achieved occupancy to assess a kernel's compute requirements, and colocates two kernels if the sum of achieved occupancy values is $<100\%$.
\footnote{\new{We focus on how Usher makes colocation decisions, not its extra components, such as operator graph merging.}}  
Achieved occupancy can be misleading, as a kernel can saturate GPU resources even with low achieved occupancy. To demonstrate this, we launch two instances of a \texttt{compute} kernel on an H100. The kernel performs several iterations of independent element-wise fp32 multiplications and we launch both instances with 132 blocks and 128 threads/block per kernel. The launch configuration is chosen to have one thread block running per SM and one warp per SMSP. NCU reports an achieved occupancy of 6.25\% per kernel, suggesting that colocation should not result in performance degradation. We show two counter-examples. First, we follow Usher's suggestion of constraining each kernel to a percentage of SMs equal to its achieved occupancy. Thus, we limit each kernel to 6.25\% of the GPU SMs, setting the \texttt{CUDA\_MPS\_ACTIVE\_THREAD\_PERCENTAGE}~\cite{mps} variable. We observe a 8.57$\times$ increase in each kernel's latency indicating that the number of SMs needed is not aligned with the achieved occupancy. Second, we run both kernels concurrently with the same launch configuration in separate CUDA streams and let them use all available SMs. We observe a 1.73$\times$ increase in each kernel's latency, despite their low achieved occupancy. These examples indicate that predicting interference and required compute resources based only on the number of active warps is not sufficient. 

Other schedulers also consider only a single metric, such as SM utilization or the utilization metric reported by nvidia-smi, which leads to suboptimal decisions~\cite{Mingcong22Reef, he2024uellm, Wu23TGS, Bhasi24serverless, Yeung20utilpred, Yeung22Horus}.

\textbf{Pitfall 2: Ignoring critical metrics.} Although GPU schedulers may consider more metrics, they often overlook essential ones. For example, Orion~\cite{Strati24Orion} colocates kernels with complementary resource profiles (i.e., compute vs. memory bound kernels), which it determines based on arithmetic intensity. However, it does not consider the IPC metric. While arithmetic intensity correlates with IPC, Orion overlooks
cases where a compute kernel’s IPC is too high and will interfere with any other colocated kernel.

To demonstrate this, we reuse the \texttt{compute} kernel from the previous example and a \texttt{copy} kernel, which repeatedly copies a 4GB input to output array.
We tune the number of iterations so the two kernels have similar execution times. 
On an H100 GPU, we launch both kernels with 132 blocks and 1024 threads/block, such that we have one thread block of each kernel running on each SM and use all available threads per SM. Using NCU, we confirm that \texttt{compute} is compute-bound and \texttt{copy} is memory-bound, thus Orion would colocate them, expecting low interference. However, we observe that the execution time of \texttt{copy} doubles under colocation. NCU shows that the \texttt{compute} kernel already saturates the available IPC per SM issuing 3.99 inst/cycle/SM on average. The colocated \texttt{copy} kernel with an IPC of 0.57 will therefore experience warp scheduling interference. \S\ref{sec:ipc-interference}  elaborates on IPC interference.

The pitfalls observed in related work raise the question: \textit{how to accurately measure GPU utilization and estimate interference?}

\section{GPU interference analysis}\label{sec:analysis}

We highlight the main GPU resources where interference can occur for popular AI workloads (e.g., LLM decode) and how to identify whether a kernel is susceptible to a specific kind of interference.

\subsection{Evaluation Setup}
\label{sec:evaluation-setup}
We develop custom CUDA benchmarks, each stressing a specific GPU resource. 
We assess the sensitivity of various workloads to resource interference by colocating them with these benchmarks. For intra-sm colocation (\S\ref{sec:intra-sm-interf}) we use CUDA streams~\cite{cuda-streams}, while for inter-SM colocation (\S\ref{sec:inter-sm-interf}), we additionally use CUDA Green Contexts~\cite{cuda-green-context} to partition SMs into mutually exclusive sets, each assigned to a separate stream.\footnote{Contrary to common belief, NVIDIA Multi-Process Service (MPS) does not enforce mutual exclusion of SMs between MPS clients when setting limits. It only restricts the maximum number of SMs available to a client~\cite{mps-no-mutual-exclusion}.}
Each benchmark maintains a constant level of interference throughout the lifetime of the colocated application. Most experiments focus on large language model (LLM) inference and the Time Between Tokens (TBT) during the decode phase. Increased TBT directly impacts user experience in applications such as chatbots. We use the Gemma3-1B-IT~\cite{gemma_2025} and Llama3.1-8B-Instruct~\cite{llama3-8b-instruct} models, both running on a modified fork of vLLM (May 2025) adapted to support kernel colocation. 
We evaluate workloads on a NVIDIA H100 \new{NVL (132 SMs, CUDA 12.9, driver 575.57.08)} and RTX3090 GPUs \new{(82 SMs, CUDA 12.6, driver 560.35.03)}. 


\subsection{Block Scheduler Interference}
\label{sec:block}

As described in \S\ref{sec:background_motiv}, the block scheduler assigns blocks to SMs as long as resource constraints are met. When resources are insufficient, blocks are serialized, increasing latency. We demonstrate this on an H100 GPU by colocating the decode phase of the Llama3.1-8B-Instruct model~\cite{llama3-8b-instruct} with a lightweight \texttt{sleep} kernel that repeatedly invokes  \texttt{nanosleep}. 
At each of the 10 decode steps, we launch the \texttt{sleep} kernel in a separate stream with an approximate sleep duration of 10~ms, using 132 thread blocks, to ensure one thread block per SM\footnote{We confirmed that each SM hosts one block of the \texttt{sleep} kernel by having each block print its SM ID~\cite{cuda-smid-register}.} and 128 threads per block (1 warp per SMSP).

\begin{figure}
    \centering
    \includegraphics[width=\columnwidth]{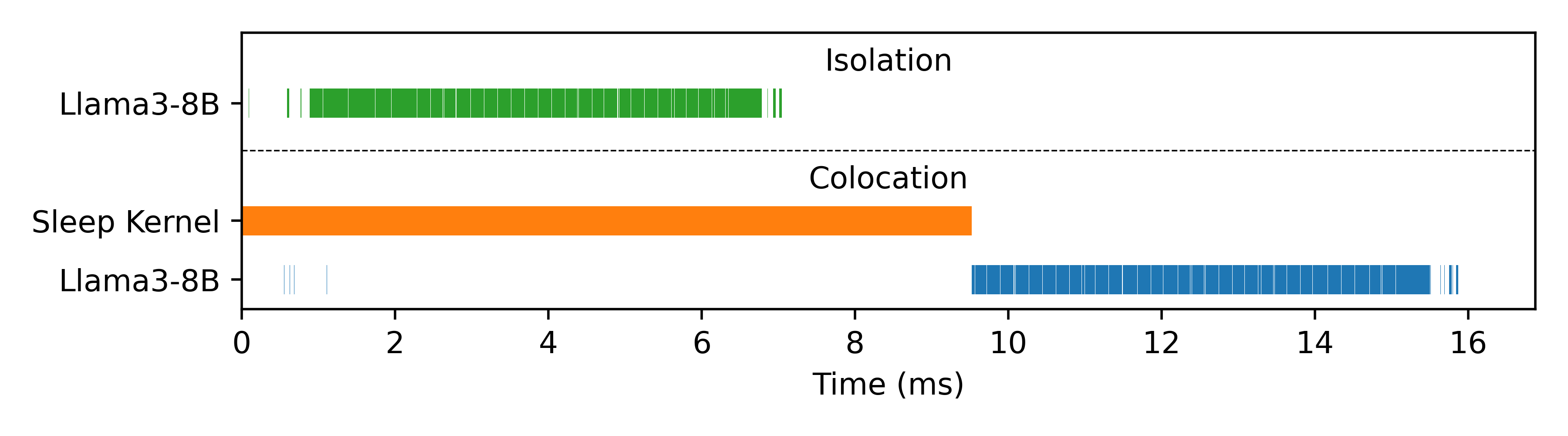}
    \caption{Decode of Llama3-8B in isolation vs colocated with a \texttt{sleep} kernel on H100 (batch size 1, prompt size 1000)}
    \label{fig:tb-scheduler-interf}
\end{figure}

In isolation, the P90 TBT of Llama3-8B (batch size 1, prompt length 1000) is 7.53 ms, increasing to 16.56 ms when colocated with the \texttt{sleep} kernel. Figure~\ref{fig:tb-scheduler-interf} shows the Nsight Systems CUDA trace~\cite{nsys} for the first decode iteration in both cases.
Although both workloads launch simultaneously, their overlap is minimal, limited to a few short kernels at the start. From the fourth kernel onward, decode execution is delayed until the \texttt{sleep} kernel completes due to SM resource contention.
The \texttt{sleep} kernel uses 16 registers per thread (2048 per block), leaving $65536-2048=63288$ registers per SM available on the H100. The fifth decode kernel, however, requires 64512 registers per block, exceeding the remaining registers and preventing concurrent execution. Consequently, the block scheduler stalls decode execution until the \texttt{sleep} kernel finishes, significanlty increasing TBT and underutilizing the GPU. Several subsequent kernels encounter the same issue. We observe similar behavior on the H100 for Gemma3-1B~\cite{gemma_2025}, though colocation with this model is feasible on the RTX3090 (\S\ref{sec:ipc-interference}).

\textbf{Takeaway:} Simply scheduling two kernels on the same GPU does not ensure colocation, as they compete for SM resources such as shared memory, registers, and threads. To prevent head-of-line blocking, where one kernel hinders others, schedulers should make decisions at fine granularity, e.g., per set of kernels~\cite{zhu2024nanoflow}, per kernel~\cite{Mingcong22Reef, Strati24Orion} or per set of thread blocks~\cite{coppock2025lithos}. The above example also shows that if developers write kernels that use as many GPU resources as possible (common today), they leave the scheduler with little opportunities to colocate and optimize efficiency.

\subsection{Inter-SM Interference: Global Memory and L2 cache}\label{sec:inter-sm-interf}

While workloads can be configured to run on separate SMs, global memory and L2 cache remain shared resources. As we illustrate below, interference can arise in two forms: (1) L2 cache pollution, which degrades locality, (2) L2 and memory bandwidth contention. 

\textbf{L2 Cache Pollution} A kernel is susceptible to L2 pollution if it repeatedly accesses data from L2 and it exhibits a high L2 hit rate. 
To study this, we colocate two instances of the \texttt{copy} kernel, in separate CUDA green contexts on the H100 (with a 50MB L2). We split the SMs across both contexts: one kernel uses 64 thread blocks (1024 threads/block) on 64 SMs; the other uses 68 thread blocks (1024 threads/block) on 68 SMs. To minimize L1 effects, we configure the kernels to maximize for shared memory. We gradually increase the data size per instance and measure the slowdown of one kernel when colocated versus running in isolation.

\begin{figure}
    \centering
    \includegraphics[trim = 0.5cm 0.4cm 0.4cm 0.5cm,clip=true,scale=0.31, width=\columnwidth]{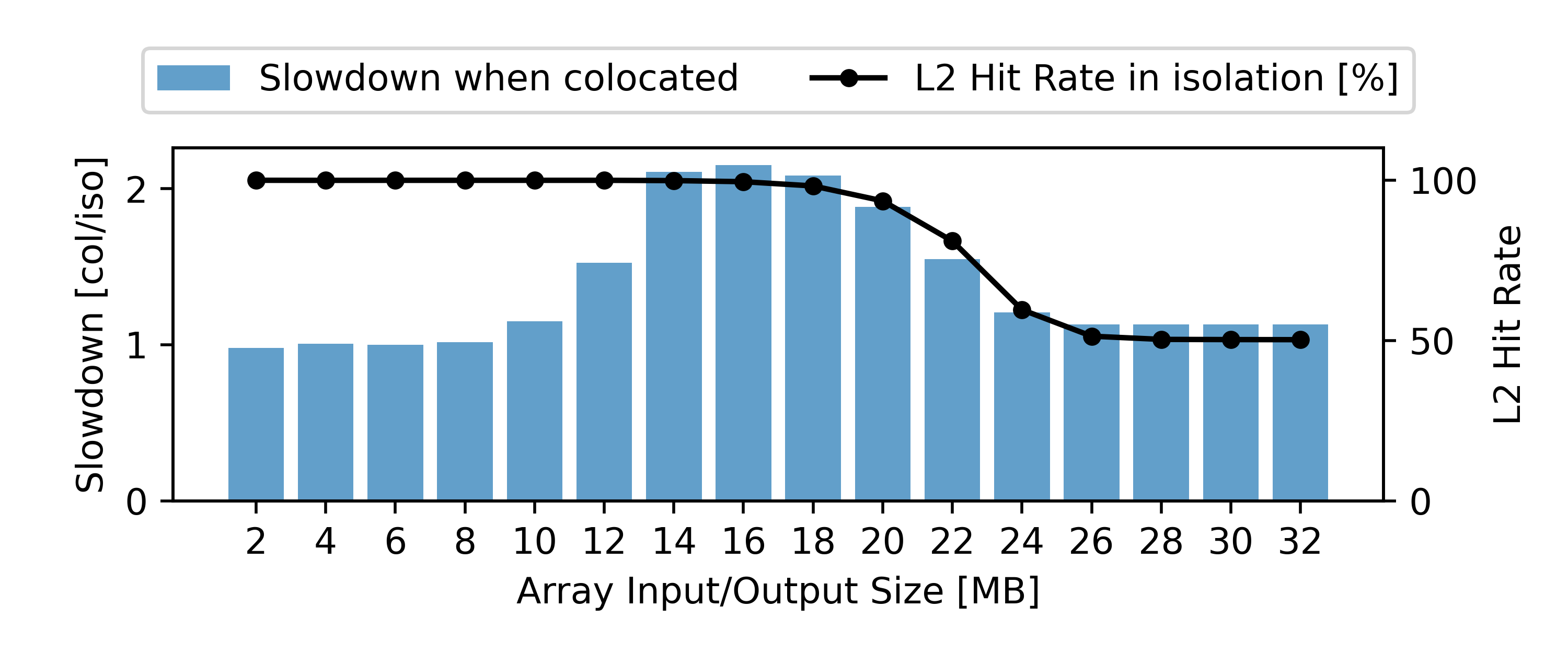}
    \caption{L2 cache interference due to cache pollution between two \texttt{copy} kernels on an H100.}
    \label{fig:l2-cache-interf}
\end{figure}

Figure~\ref{fig:l2-cache-interf} shows the results along with the isolated L2 hit rate. 
For input sizes $\leq$ 8MB, there is no slowdown, as both instances' input and output arrays fit into L2.  At 16MB, slowdown peaks at 2.15$\times$ due to the combined 64MB working set exceeding L2 capacity. 
Beyond 26MB, slowdown plateaus around 1.12$\times$, as L2 locality in isolation is lost (more data needs to be loaded from main memory). Further L2 pollution by a colocated kernel creates no additional slowdown. The L2 hit rate stabilizes near 50\%, as every store is counted as a hit by NCU, making half of accesses appear as hits~\cite{ncu-l2-hit-rate-stores}. 

\textbf{L2 Cache/Memory Bandwidth Interference}
Kernels with high memory demand often lack L2 locality, making them less prone to pollution but more sensitive to bandwidth contention. To demonstrate this, we colocate the decode phase of Llama3.1-8B with a \texttt{copy} kernel that transfers a 4GB input using vectorized loads. Each workload runs in its own CUDA green context, with 64 SMs allocated to Llama and 68 SMs to the \texttt{copy} kernel.
Table~\ref{tab:membw-interference} reports the P90 TBT when generating 10 tokens for the LLM running alone vs. when colocated with the \texttt{copy} kernel, along with the copy kernel's L2 and Memory bandwidth throughput utilization.
As the number of \texttt{copy} thread blocks increases, bandwidth pressure rises, causing up to 1.3$\times$ slowdown. 
Since L2 access is mandatory, higher memory bandwidth also translates to higher L2 throughput. We observe similar trends across other prompt and batch sizes. We observe that all LLM decode kernels 
do not benefit much from L2 cache locality and have to load most of their data from main memory. 
This makes them susceptible to L2 and memory bandwidth interference; in the examples of Table ~\ref{tab:membw-interference}, most LLM decode kernels experienced significant slowdown.

\textbf{Takeaway:} Even when workloads run on disjoint SMs, using strict SM isolation mechanisms such as green contexts, they can still contend for L2 capacity and L2 and global memory bandwidth.

\begin{table}
\centering
\footnotesize 
\begin{tabular}{lccccc}
\toprule
\textbf{Thread Blocks of interference kernel} & \textbf{0} & \textbf{34} & \textbf{68} & \textbf{102} & \textbf{136} \\
\midrule
\textbf{L2 Band Util (copy kernel) [\%]}  &  & 37 & 68 & 87 & 95 \\
\textbf{Mem Band Util (copy kernel) [\%]}  &  & 27 & 51 & 69 & 81 \\
\textbf{P90 TBT (ms)} & 16.9 & 17.6 & 18.38 & 19.92 & 22  \\
\bottomrule
\end{tabular}
\caption{Effect of memory bandwidth interference on the Time Between Tokens for the LLAMA-3.1-8B on H100 GPU, batch size 8, 16384 tokens per prompt. When the LLM runs on the whole GPU (no green contexts), P90 TBT is 14.2 ms}
\label{tab:membw-interference}
\vspace{-0.5cm}
\end{table}

\subsection{Intra-SM Interference}
\label{sec:intra-sm-interf}
When workloads run on the same SM, they can interfere in the warp scheduler, the computation pipelines or shared memory.

\subsubsection{Shared Memory}
\label{sec:shared-memory}
As mentioned in \S\ref{sec:background:gpu-arch}, each SM has a private memory space partitioned between the L1 cache and shared memory. 
To achieve high performance, kernels with high arithmetic intensity load chunks of data from global memory into shared memory for computation before writing results back to global memory. 
Modern NVIDIA GPUs ($\geq$ CC 5.x) have 32 memory banks, each with a bandwidth of 32 bits/cycle. Successive 32-bit words map to successive banks. 
An \textit{n-way bank conflict} occurs when $n>1$ threads within the same warp access different data in the same bank.
An $n$-way bank conflict is resolved by serializing the requests over $n$ conflict-free requests, thereby decreasing throughput. 

\begin{figure}
    \centering
    \includegraphics[trim = 0.5cm 0.4cm 0.4cm 0.5cm,clip=true,scale=0.31, width=\columnwidth]{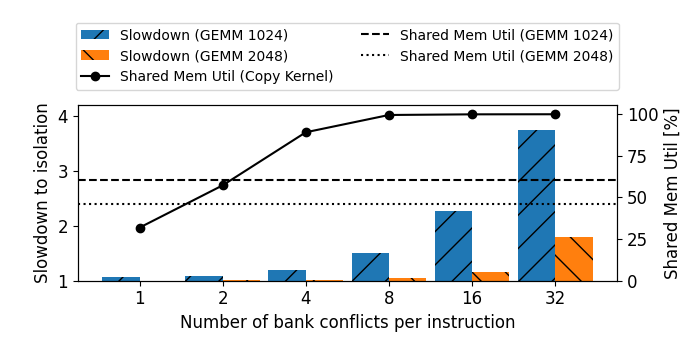}
    \caption{Shared Memory interference. Effect of increasing shared memory bank conflicts on two cuBLAS GEMM implementations for dimensions 1024 and 2048 on an H100.}
    \label{fig:shared-memory-interf}
    \vspace{-0.5cm}
\end{figure}

Kernels running concurrently on the same SM can contend for shared memory bandwidth.
We demonstrate this by colocating the PyTorch \texttt{torch.mm} operator~\cite{torch_mm} with a custom $n$-strided copy kernel that repeatedly loads and stores elements of a 4KB array within shared memory. 
Varying the access stride affects the shared memory bank conflicts within a warp.
Figure~\ref{fig:shared-memory-interf} shows the slowdown under colocation vs. running in isolation of the \texttt{torch.mm} operator on the H100 as we increase the number of bank conflicts within our copy kernel. 
Depending on the input matrix dimensions, \texttt{torch.mm}, uses different implementations of cuBLAS GEMM. 
For a 32-way bank conflict, the slowdown is 1.79$\times$ for dims 2048 and 3.75$\times$ for dims 1024. 
The cuBLAS implementation for dimension 1024 has a higher utilization of the shared memory pipeline, which explains its higher sensitivity to interference. 
The slowdown for both implementations increases significantly beyond 4 bank conflicts. Since the shared memory pipeline is already saturated by the \texttt{copy} kernel, further increases in bank conflicts lead to further serialization of shared memory requests, thereby increasing the \texttt{torch.mm} latency.

\textbf{Takeaway:} A kernel with non-optimal shared memory access may result in many bank conflicts, saturating the SM’s shared memory pipeline and starving memory accesses of colocated kernels. 
\subsubsection{IPC interference}
\label{sec:ipc-interference}

To demonstrate how the architectural limit of 4 instr/cycle/SM can become a bottleneck, we colocate the memory-bound decode phase of the Gemma3-1B model~\cite{gemma_2025} with four versions of our \texttt{compute} kernel ($S_1$ to $S_4$), each increasing the Instruction Level Parallelism (ILP) to put growing pressure on the warp scheduler. 
To enable concurrent intra-SM execution and avoid sequential kernel launches due to shared memory constraints, we configure the L1 cache/shared memory split to maximize shared memory~\cite{cuda-shared-mem-conf}. We conducted the experiments on the RTX3090 GPU, as certain Gemma kernels on the H100 saturate the thread capacity per SM, preventing intra-SM colocation (\S\ref{sec:block}).
In each scenario ($S_0$ to $S_4$), we generate 10 new tokens and report the 90th percentile of TBT. 
For scenarios $S_1$ to $S_4$, we launch the modified \texttt{compute} kernel with 82 thread blocks (one per SM) and 128 threads per block (one warp per SMSP).
Scenario $S_0$ represents the baseline, with the Gemma model running in isolation and without setting any custom L1 cache/shared memory configuration.

\begin{figure}
    \centering
    \includegraphics[trim = 0.4cm 0.4cm 0.4cm 0.5cm,clip=true,scale=0.31, width=\columnwidth]{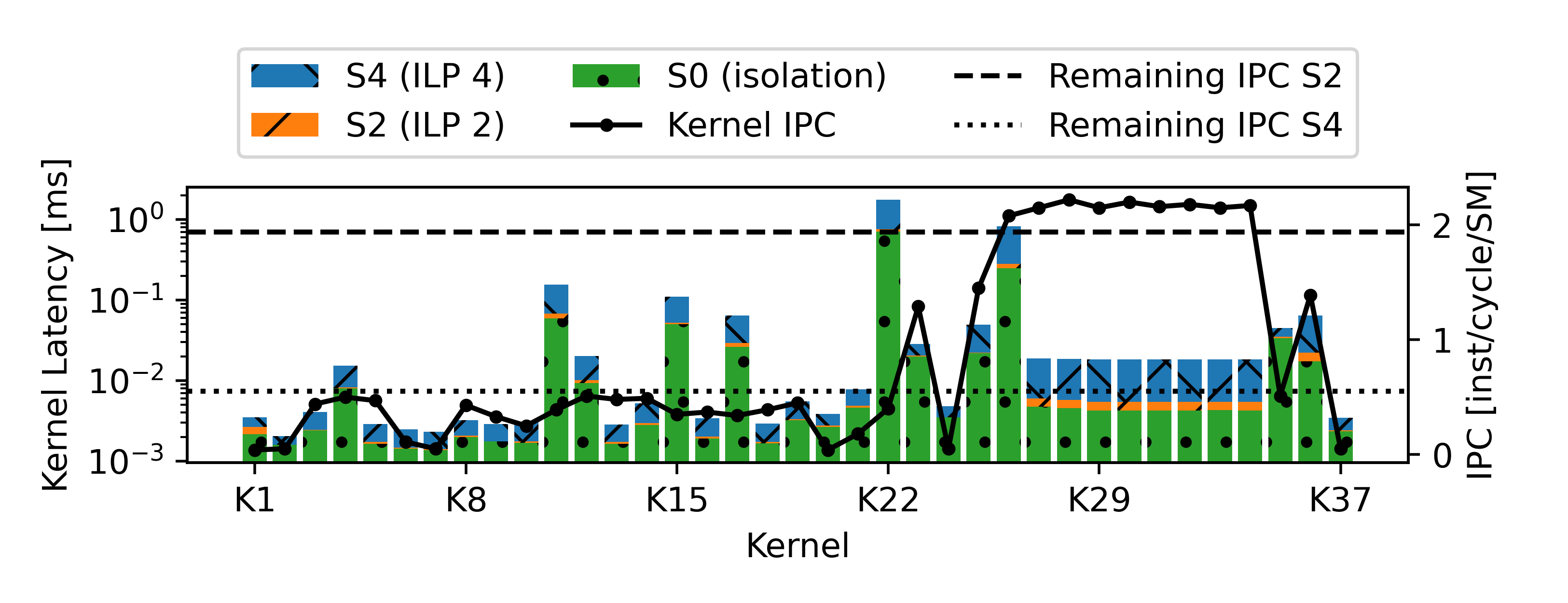}
    \caption{Kernel latencies for scenarios $S_0, S_2, S_4$ running a Gemma3-1B decode iteration with a single hidden layer on the RTX3090 (batch size 8, prompt size 1000). Dashed lines represent left-over IPC when \texttt{compute} kernel is running.
    }
    \label{fig:ipc-decode-lats}
\end{figure}

Table~\ref{tab:ipc-interference} shows that the TBT remains largely unaffected in scenarios $S_0$ to $S_3$, but experiences significant degradation as the \texttt{compute} kernel’s IPC approaches the hardware limit in $S_4$. 
Comparing $S_0$ and $S_1$ shows that manually setting the shared memory configuration to maximize for shared memory has no noticeable impact on the TBT of the Gemma model. Nevertheless, this potential impact should be kept in mind whenever manually setting the configuration.
Figure~\ref{fig:ipc-decode-lats} highlights IPC interference for a prompt of 1000 tokens and batch size 8, illustrating the latency slowdown experienced by each kernel for the Gemma3-1B model configured with a single hidden layer (in contrast to 26 in the full model). 
In $S_2$, only kernels $K26$ to $K34$ experience high slowdown due to high IPCs. However, in $S_4$, all kernels slow down as the combined IPC approaches or exceeds the architectural limit of 4~\textit{instr}/cycle/SM. This highlights that intra-SM colocation under limited interference is feasible, but must be managed carefully to prevent performance degradation.

\begin{table}
\centering
\footnotesize 
\begin{tabular}{lccccc}
\toprule
& \textbf{$S_0$} & \textbf{$S_1$} & \textbf{$S_2$} & \textbf{$S_3$} & \textbf{$S_4$} \\
\midrule
\textbf{IPC \texttt{compute}} [instr/cycle/SM] & 0 & 1.18 & 2.06 & 2.9 & 3.45 \\
\midrule
\textbf{Prompt Size 1000, Batch Size 1} & 5.59 & 5.75 & 6.00 & 6.24 & 10.74 \\
\textbf{Prompt Size 1000, Batch Size 8} & 6.08 & 6.23 & 6.56 & 6.94 & 12.52 \\
\bottomrule
\end{tabular}
\caption{P90 Time Between Tokens (TBT) latency in \texttt{ms} when generating 10 tokens with the Gemma3-1B~\cite{gemma_2025} model on the RTX3090. The model is run in isolation ($S_0$) and colocated with a compute bound kernel that emits an increasing number of instructions per cycle ($S_1-S_4$).}
\label{tab:ipc-interference}
\vspace{-0.75cm}
\end{table}

\textbf{Takeaway:} While blocks from different kernels can benefit from colocation on the same SM, a kernel that has a very high IPC can cause significant slowdown to other kernels, due to warp scheduling interference. In that case, scheduling to separate SMs (e.g., using green contexts) might be a better approach.

\subsubsection{Pipeline Interference}
\label{sec:pipeline-interf}
We modify our \texttt{compute} kernel to utilize fp64 multiplication \texttt{\_\_dmul\_rn}~\cite{cuda-fp64-intrinsics}. We examine scenarios $S_1$ to $S_4$ increasing ILP from one to four. We colocate two kernels, each with 132 thread blocks and 128 threads per block, to have one thread block per SM and one warp per kernel on each SMSP.
Table \ref{tab:pipeline-interference} shows that when the \textit{aggregated} FP64 pipeline utilization (computed by summing up the per-kernel utilizations from Table~\ref{tab:pipeline-interference}) remains below 100\% ($S_1$ and $S_2$), colocation offers a significant speedup ($>1.87\times$). 
However, the speedup decreases significantly as utilization exceeds 100\% ($S_3$ and $S_4$), despite IPC not being a bottleneck. 
Even though saturating a computation pipeline is frequently associated with a high IPC, compute interference can still occur before the warp scheduler saturates. In the H100 architecture, the peak performance for arithmetic FP64 operations is only half that of FP32 operations, thus we demonstrate interference for FP64 operations.

This example also illustrates why the achieved occupancy alone is insufficient for colocation decisions, as already seen in \S\ref{sec:issues-of-existing-estimations}. 
Despite having an achieved occupancy of 6.25\% in all four scenarios, the kernel saturates an entire computation pipeline due to a high number of independent arithmetic instructions. While this scenario is less likely to occur in practice (since most GPU workloads do not use FP64), this demonstrates another potential source of interference.

\begin{table}
\centering
\footnotesize 
\begin{tabular}{lcccc}
\toprule
& \textbf{$S_1$} & \textbf{$S_2$} & \textbf{$S_3$} & \textbf{$S_4$} \\
\midrule
\textbf{IPC \texttt{compute}} [instr/cycle/SM] & 0.64 & 1.10 & 1.53 & 1.96 \\
\textbf{FP64 Pipe Utilization} [\%] & 24.22 & 47.71 & 69.42 & 90.68 \\
\textbf{Speedup} [seq/col] & 1.93 & 1.87 & 1.33 & 1.03 \\
\bottomrule
\end{tabular}
\caption{Speedup of colocating two FP64 \texttt{compute} kernels over running them sequentially on H100. IPC and FP64 Pipe Util correspond to profiling the \texttt{compute} kernel in isolation.}
\label{tab:pipeline-interference}
\vspace{-0.75cm}
\end{table}

\textbf{Takeaway:} SM's pipelines can saturate even before the warp scheduler saturates, which might require scheduling to different SMs to avoid pipeline interference.

\section{Research Outlook and Conclusions}

We discuss how insights from \S\ref{sec:analysis} can guide the design of an interference-aware GPU scheduler (\S\ref{sec:discussion-scheduling}), how GPU vendors can enable more flexible and efficient colocation (\S\ref{sec:discussion-vendors}), and discuss how GPU programmers can develop GPU kernels with colocation in mind (\S\ref{sec:discussion-design}).

\subsection{Interference-aware GPU Scheduling}\label{sec:discussion-scheduling}

Our analysis in \S\ref{sec:analysis} helps us identify key requirements that a GPU scheduler should satisfy to enable efficient GPU workload colocation with strict performance guarantees. First, scheduling decisions should be made at a \textit{fine granularity}, per set of kernels~\cite{zhu2024nanoflow}, per kernel~\cite{Strati24Orion}, or per thread blocks~\cite{coppock2025lithos}, to avoid head-of-line blocking caused by long-running, resource-demanding kernels preventing colocation, as shown in \S\ref{sec:block}.

Second, a scheduler should be able to predict and quantify interference between colocated kernels and workloads. As a first step, we propose developing a kernel-level
interference estimator to predict the performance of kernels under colocation. For each workload, the estimator can collect each
kernel’s metrics and resource sensitivity as outlined in \S\ref{sec:analysis}. The
estimator can then predict each kernel’s slowdown due to interference at each resource. Existing interference estimators only consider a subset of interference sources~\cite{Xu2022iGniter, Zhao20HSM}. Themis~\cite{Zhao19Themis} and GPUPool~\cite{Tan22GPUPool} consider many of the resources outlined in \S\ref{sec:analysis}, but treat them as a black-box input to an ML model, and present their analysis and evaluation only in simulation. Instead, we demonstrated interference caused by contention for these resources on high-end NVIDIA GPUs. The kernel-level estimator provides a foundation for implementing a workload-level interference estimator that can predict a job's interference sensitivity. Using that estimator, a GPU scheduler can find workload pairs suitable for colocation.

Third, a GPU scheduler should balance per-workload performance and GPU efficiency. For example, restricting a workload or a set of kernels to a subset of SMs can increase GPU efficiency with acceptable performance degradation (in \S\ref{sec:inter-sm-interf}, we saw that restricting the LLM decode to less than half of the GPU's SMs has only a 1.19$\times$ TBT slowdown). Adjusting SM L1/shared memory configurations can also improve colocation, as we saw in \S\ref{sec:ipc-interference}. 

Finally, GPU schedulers should account for  differences in GPU generations.  Libraries such as cuDNN and cuBLAS have different implementations for different GPU architectures, meaning that the same workload might have different behavior on two different GPUs, as shown for the LLM decode phase in \S\ref{sec:analysis}. Thus, a scheduler should reprofile a GPU workload when running on a different GPU. \new{Although profiling with tools such as NCU can span minutes to hours, this is acceptable since models are trained or deployed for large periods of time.}

\subsection{Hardware Support for Spatial GPU Sharing}\label{sec:discussion-vendors}

The closed-source nature of NVIDIA GPUs limits
user control over kernel execution as various hardware mechanisms
are a black box. Exposing some hardware features can help programmers take better control of the GPU. Better intra-SM visibility
is needed, providing insights into the warp scheduling algorithm
and the mapping of instructions to physical cores. Enhancing profiling tools such as NCU to allow for collecting GPU metrics under kernel colocation would be very beneficial for better understanding interference, although very challenging since these profilers heavily rely on deterministic kernel replay. Furthermore, programmer-friendly ways to partition SMs and DRAM channels
at the kernel level can mitigate intra-SM and DRAM bandwidth
interference. CUDA green contexts already provide much stricter isolation compared to MPS, but they only partition SMs and not DRAM channels.  Related work proposes limiting a kernel's blocks to specific DRAM channels, but they are code-intrusive and unsuitable for ML workloads with closed-source kernels~\cite{zhang2024missile, Jain19fgpus, Zhao23Miriam, Chen23Combo}.  Finally, enabling kernel preemptibility could improve kernel colocation, especially in real-time tasks, as shown by REEF~\cite{Mingcong22Reef} for AMD GPUs.

\subsection{GPU Kernel Design Tradeoffs} \label{sec:discussion-design}

Most existing high-performance GPU workloads are not designed with colocation in mind. Kernels often will try to maximize their intra-SM resource usage (registers, amount of shared memory), and launch large grid sizes, preventing the thread block scheduler from running anything else in the GPU at the same time (see \S\ref{sec:block}). However, as related work has shown, even high-performance kernels leave parts of the GPU severely underutilized~\cite{Strati24Orion, kamath2024podattention}. This presents a tradeoff between maximizing individual kernel performance and enabling efficient colocation~\cite{zhu2024nanoflow}.
Marginal performance gains may come at the cost of GPU utilization and overall efficiency. Thus, if the goal is to increase GPU efficiency with workload colocation, GPU programmers can aid the GPU schedulers by providing colocation-friendly kernel variants.


We hope our characterization of GPU interference inspires and informs future work on GPU scheduling and custom kernel design, \new{and we encourage similar studies across other vendors (e.g. AMD).}

\section{Acknowledgments}
\new{We thank the HotOS'25 and SoCC'25 reviewers for their insightful feedback. Foteini Strati is supported by the Swiss National Science Foundation (project number 200021\_204620). Natalie Enright Jerger is supported in part by the Canada Research Chairs program.}

\bibliographystyle{plain}
\balance
\bibliography{references}

\end{document}